\documentclass[reqno]{amsart}
\usepackage[utf8]{inputenc}
\usepackage{amscd,amssymb,amsmath,amsthm}
\usepackage{amsmath,systeme}
\usepackage{psfrag}
\usepackage{color}
\usepackage{hyperref}
\usepackage{placeins}
\usepackage{xcolor}
\hypersetup{
    colorlinks=true,
    linkcolor=blue,
    filecolor=magenta,      
    urlcolor=cyan,
    pdftitle={Overleaf Example},
    pdfpagemode=FullScreen,
    }
\usepackage{caption,subcaption,comment}
\usepackage[percent]{overpic}
\usepackage[centering,marginparwidth=2cm]{geometry}
\usepackage{marginnote}
\def\bf{\mathbf{f}}

\def\bR{\mathbf{R}}

\def\eps{\varepsilon}

\theoremstyle{definition}

\def\beq{\begin{equation}}
\def\eeq{\end{equation}}

\numberwithin{equation}{section}

\begin{document}

\title[A Modified PINN Approach for Compartmental Models in Epidemiology]{A Modified PINN Approach for Identifiable Compartmental \\ Models in Epidemiology with Applications to COVID-19}
\author{Haoran Hu}
\address{Department of Mathematics and Statistics, UMass Amherst, Amherst, MA 01003}
\email{haoranhu@umass.edu}
\author{Connor M. Kennedy}
\address{Department of Mathematics and Statistics, UMass Amherst, Amherst, MA 01003}
\email{conmkennedy@umass.edu}
\author{Panayotis G. Kevrekidis}
\address{Department of Mathematics and Statistics, UMass Amherst, Amherst, MA 01003}
\email{kevrekid@umass.edu}
\author{Hong-Kun Zhang}
\address{Department of Mathematics and Statistics, UMass Amherst, Amherst, MA 01003}
\email{hongkun@math.umass.edu}

\subjclass[2000]{37D30}

\keywords{Network dynamics, Covid-19, PINNs, Wavelets}

\begin{abstract}
A variety of approaches using compartmental models have been used to study the COVID-19 pandemic and the usage of machine learning methods with these models has had particularly notable success. 
We present here an approach toward analyzing accessible data on COVID-19's U.S. development using a variation of the “Physics Informed Neural Networks” (PINN) which is capable of using the knowledge of the model to aid learning. We illustrate the challenges of using the standard PINN approach, then how with appropriate and novel modifications to the loss function the network can perform well even in our case of incomplete information.
Aspects of identifiability of the model parameters are also assessed, as well as methods of denoising available data using a wavelet transform. Finally, we discuss the capability of the neural network methodology to work with models of varying parameter values, as well as a concrete application in estimating how effectively cases are being tested for in a population, providing a ranking of U.S. states by means of their respective testing.
\end{abstract}

\maketitle

\section{Introduction}

On December 31st, 2019, 27 cases of pneumonia were reported in Wuhan City, Hubei Province in China. The cause was identified on January 7th, 2020 and was subsequently termed SARS-CoV-2 (the virus) and COVID-19 (the disease) by the World Health Organization (WHO) \cite{WHO_Review}. The disease subsequently grew to the point that the WHO officially declared it a pandemic on March 11, 2020 \cite{WHO_Pandemic_Declaration}. As of May 31st, 2022, a cumulative total of 526,558,033 cases and 6,287,117 deaths attributed to the disease have occurred \cite{WHO_COVID_stats}. The sheer scale of the disease's development has prompted research to combat it across the planet. Proper modeling of the disease has been crucial in not only evaluating the virus' behavior itself, but the effects of policy decisions made in response to it~\cite{review_meta,holmdahl2020}. Extensive evaluation of social-distancing measures and other non-pharmaceutical interventions found that these approaches were effective, supporting the adoption of these policies 
on numerous occasions throughout the pandemic~\cite{non-pharma_intervention,galvani2020,chinazzi2020}. More recently, the 
role of vaccination has been also considered in the relevant
models~\cite{vacc2021}.

One of the most critical factors in studying the disease has been the difficulty in gathering information about its spread. Even simply getting an accurate estimate of the current number of infected individuals has been extremely difficult~\cite{bhatta2020}. Accurate counts of the infected population are crucial, as a guiding tool on policy decisions from the distribution of testing resources, allocation of treatment materials, and the severity of lockdown procedures.  Limitations on testing supplies in the early stages, and the frequency of asymptomatic cases, have created major information gathering difficulties \cite{Asymptomatic}. Accurate estimation of certain parameters used in modeling the disease, notably the base reproduction rate and case confirmation rate, are also useful in guiding policy but their true values are dependent on these unknown cases. There are some methods however, to infer these unknown quantities from limited known data; see, e.g.,~\cite{george2022} for a recent discussion. We focus here specifically on the usage of compartmental epidemiological models in conjunction with the usage of the widely applicable and highly successful technology of the so-called Physics Informed Neural Networks (PINNs)
(see~\cite{karnia} for a recent review thereof) to estimate these quantities.

The classic SIR ODE model is, arguably, the most well-known one, separating the population into susceptible, infected, and removed
(recovered) groups, going back to the seminal work of \cite{Math_Epidemic}. In the almost century that has followed there has been a wide variety of approaches to and variations of the model to fit the particular features of different diseases \cite{Epidemic_Summary}. In the case of COVID-19, several modifications have been used to more accurately describe its development.The SEAHIR model (involving susceptible, exposed, asymptomatic, hospitalized, infected and recovered populations), incorporates the effect of social isolation measures \cite{SEAHIR}, while the SIRSi (Susceptible, Infected, Recovered, Sick) model represents individuals that lose immunity after recovering from an infection \cite{SIRSi}. 
There have been numerous other proposals, involving different
numbers of components (and other features such as, e.g., age stratification~\cite{cuevas2021} and spatial distribution~\cite{arenas2020}).
More complex models do run the risk of their unknown quantities being difficult to estimate, or in some cases even impossible \cite{Practical}. If we want to estimate these unknowns, we need a model that reflects the practical realities of COVID-19, the effects of isolation~\cite{non-pharma_intervention} and the prevalence of asymptomatic cases~\cite{Asymptomatic,bhatta2020}, while still allowing estimations. 

We adopt the usage of the SICRD model implemented in \cite{ZhangDong20}.
While the susceptible, infected and recovered groups are similar
to the standard SIR setting discussed above, the two additional
groups added herein concern
the confirmed cases, and the deaths. We adopt this model as it incorporates the noteworthy effects of testing and quarantining,
and crucially for our purposes, it splits the infected population into unknown/known compartments, as well as includes the information on death data. The latter is an extremely important piece of accessible information as we note in section \ref{Numerical Results on Identifiability}. Indeed, without it, it would actually be impossible to estimate the number of infected individuals or most of the important parameters from confirmed case counts alone. 
The relevance of the inclusion of the death data 
(as well as potentially and more recently, the data on
hospitalizations) is a feature that some of the present
authors have argued in various earlier works~\cite{cuevas2021,george2022}.
The SEAHIR model has similar advantages and much of
the analysis below could be adapted to that model as well (an
avenue that we do not pursue herein).

We propose the usage of a sophisticated neural network approach to estimate the unknown cases and parameters, i.e., a methodology that has proved extremely successful in a variety of applications within scientific computing~\cite{karnia,PINN_Survey}. There have been promising recent results along this vein through the usage of neural networks for forecasting the development of COVID-19 in Germany~\cite{GNN_Germany} and the spread of influenza in the U.S.~\cite{Multires_NN}. These make use of intensely studied
recent  tools within machine learning, such as graph neural networks and recurrent neural networks, both of which have structures adapted to better suit the features of the problem. Our model is particularly well suited for an adaptation of the ``Physics Informed Neural Network" (PINN) approach to our problem~\cite{PINN}. 

The principle of PINNs was developed for the estimation of unknown quantities in a system adhering to a physical law, generally a nonlinear PDE (or lattice dynamical equation~\cite{weizhu}) that the system is assumed to obey \cite{PINN}. The effectiveness of this approach has been substantial across a wide variety of disciplines, notably having excellent compatibility with other techniques such as the previously mentioned graph and recurrent neural networks \cite{PINN_Survey}. Some adaptation of this approach to disease modeling has already been attempted \cite{DINN}, but we expand on this idea with a novel approach to the training loss. We find that in our case where several population compartments are not directly known that the standard PINN loss function is insufficient. We derive a new set of terms for our loss function in section \ref{Setup of the Neural Network}. With several key simplifications in the loss terms we substantially improve the stability and efficiency of the network's performance. The efficacy of our results, even with a simple feed-forward network architecture, suggests the relevance of further research into combining with recurrent networks to improve performance, or application to a graph neural network to incorporate human mobility~\cite{arenas2020,PNAS}.

We begin in section \ref{The SICRD Model} by explicitly defining the SICRD model and the parameters within, 
as well as explaining the general concepts of the model. With the model defined, in section \ref{Identifiability} we present the  definition of identifiability, the concept of when parameters may be uniquely determined by the known variables in a system. The identifiability of our model is verified, and the lack thereof for several slight variations is shown, which justifies our particular choice. In section \ref{Denoising of data using wavelet transform} we address the issue of potential noise in our collected data with the use of a wavelet transform, which separates the signal into the key features and the noisy component caused by smaller scale fluctuations. This filtering via the transform also has the added benefit of smoothing the data, which improves our ability to apply an ODE model to it. With the model defined and the data appropriately processed, section \ref{Setup of the Neural Network} explicitly defines our specific selection of the loss function for the network along with corresponding explanation for its usage.

We first demonstrate the neural network's effectiveness by running estimations on the ideal case, with simulated data that explicitly obeys our model in section \ref{Artificial Data Testing}. In section \ref{time dependent parameters} we also demonstrate that, due to the high efficiency in the network's calculations, it may also be used to perform estimates using a model with time-varying parameters as opposed to simply assuming constant parameters, and no particular restrictions on the form of time-dependence are necessary. Our results show that when the parameters are assumed to vary, the variance is substantial enough that over any longer time frame  constant parameter assumptions would be highly unrealistic. 
We demonstrate the network's ability to estimate the unknown infected population given real data reported in U.S. states in section \ref{real data testing}. 

In section \ref{Ranking States by Testing} we propose a ranking of how well the states are conducting testing by calculating the ratio of the (predicted as) infected population over the positively tested population. The intention is to give a more precise estimate on how effectively testing is being conducted in a region, rather than simply looking at the per capita number of infections. This can highlight population centers where infections are low, as the disease has only entered it recently, but the infrastructure for conducting testing is poor. If a region with poor testing can be recognized early, then the problem can be addressed before it reaches a major outbreak. 

Of note for all our results is that our methods applied herein
could easily be adapted to other countries or regions, as well as considering population centers of smaller scales. Our modified PINN approach is also compatible with a variety of other effective techniques in machine learning and has the possibility for extensive further research.

\section{Data, Model and method}

\subsection{The SICRD Model}\label{The SICRD Model}

\begin{figure}[ht]
  \hspace{-0.5in}
    \includegraphics[width=5in]{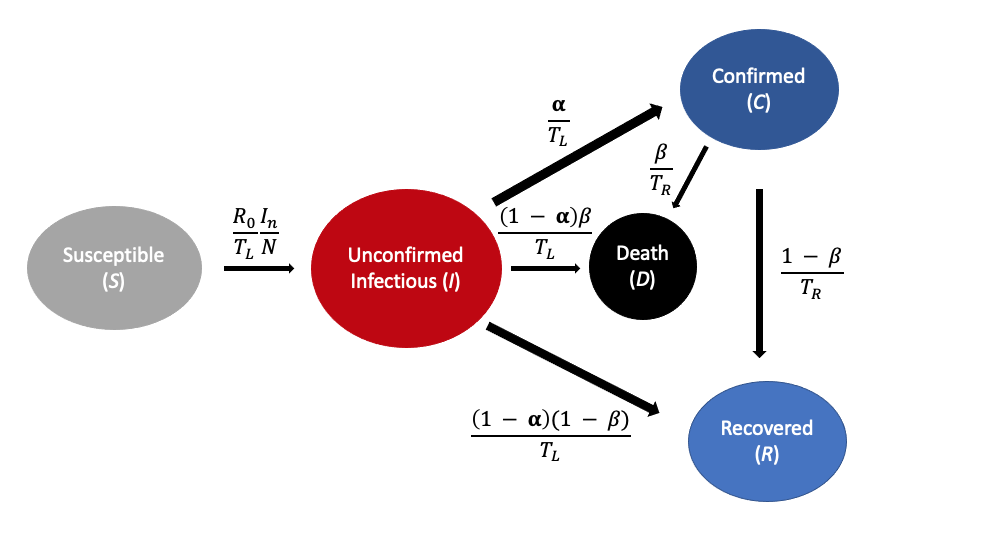}
    \caption{Schematic of the SICRD model inspired from~\cite{ZhangDong20} illustrating the main populations and interactions.}
    \label{fig:schematic}
\end{figure}

In modeling the development of COVID-19, we focus on U.S. states as our population centers due to the relative ease of data access. In each state, we use a modification of the SIR model \cite{ZhangDong20} by introducing to the traditional SIR model the populations referring to the number of deaths and that of the confirmed cases. The SICRD model is a compartmental model in which the population is divided into susceptible (S), infectious (I), confirmed (C),  recovered (R) and dead (D) individuals; 
see also Fig.~\ref{fig:schematic}. Note that each compartment refers to the current count and not the total cumulative number of cases. 

This model reflects the real-life situation in which an infectious person may recover before receiving a formal diagnosis, as well as accounting for the effect of testing for the disease and quarantining. The model is defined explicitly by the following ODE system:
\begin{eqnarray}\label{SICRD Unscaled I to D}
\dot S&=& -\frac{R_0}{T_L} \frac{S I}{N} \\
\dot I&=& \frac{R_0}{T_L} \frac{S I}{N} -\frac{1}{T_L} I \nonumber\\
\dot C&=& \frac{\alpha}{T_L} I - \frac{1}{T_R} C \nonumber \\
\dot D&=& \frac{\beta(1-\alpha)}{T_L} I + \frac{\beta}{T_R} C \nonumber\\
\dot R&=& \frac{(1 - \beta)(1 - \alpha)}{T_L} I + \frac{1 - \beta}{T_R}C, \nonumber
\end{eqnarray}
where $N=S+I+C+D+R$ is the total population size. $R_0$ is the basic reproductive number, which refers to the the average number of cases directly infected by one infectious case in a completely susceptible population. $T_L$ is the average number of days from beginning to be infectious to ceasing to be infectious.
$T_R$ is the average number of days from being confirmed to recovery or death. $\alpha$ is the proportion of confirmed cases among all cases transferred from the unconfirmed infectious state. Finally, $\beta$ is the  fatality rate. 
The deaths ($D$) and recoveries ($R$) stem from either the infectious compartment ($I$) 
or from the confirmed infection one ($C$). We also note we assume individuals who have been confirmed to be infectious sufficiently quarantine and do not infect any further individuals.
Our goal is to estimate the unknown infected population $I(t)$, along with the unknown parameters $\alpha$, $\beta$, and $R_0$. We wish to perform estimation using only our accessible data, the count of confirmed cases $C(t)$, deaths $D(t)$, and the more directly estimable time scale parameters $T_L, T_R$.

The reasons for this precise formulation of the SICRD model are discussed further in section (\ref{Identifiability}). We note that other, much more complex compartmental epidemiological models with more compartments such as those presented in \cite{cuevas2021} have also been considered; here, inspired
by the latter work, we present a reduced version of the model for reasons also relating to identifiability, as discussed below. Furthermore, this relatively simple model was chosen to test the capabilities of our neural network approach without added complications, while testing of more detailed models is 
left to be considered in future work.

\subsection{Data Set}
All of the actual data on COVID-19's development used in this paper was pulled from the tool developed in \cite{covid_data}. The tool aggregates data from a variety of data sources including the WHO, each state's individual department of health, and the CDC. The data used includes reports on case counts, active cases, and deaths within each individual U.S. state. The usage of this data is in sections \ref{Denoising of data using wavelet transform}, \ref{time dependent parameters} and \ref{Ranking States by Testing}.

\section{Identifiability}\label{Identifiability}

\subsection{Identifiability Definitions}

In order to reasonably use (\ref{SICRD Unscaled I to D}) we need to verify the identifiability of the model. Identifiability is the ability to uniquely identify the model parameters from the known variables. We know from the results of~\cite{Waterborne,cuevas2021,Practical} that many compartmental epidemiological models have potential issues with identifiability. To be concrete, we recall the precise definition of the term. 

Consider an $n$-dimensional ODE system. We let $p\in \Omega_p \subset \mathbb{R}^{n_p}$ be the vector of constant parameters for the ODE system where $\Omega_p$ is our allowable parameter space. We note that the initial values of the variables in the system may also be considered parameters. We let $m(t,p)\in\mathbb{R}^{n_m}$, $h(t,p) \in\mathbb{R}^{n_h}$ be the functions representing all known and unknown variables of the ODE system respectively, with $n_m + n_h = n$. Then, the ODE system can be represented by \cite{Practical}:
\begin{align}
    \dot{m} = f(m, h ,p) \\
    \dot{h} = g(m, h ,p), \nonumber
\end{align}
Note that the same ODE may be treated with different separations into $m$ and $h$ depending on what variables are assumed known. Given such a system and a choice of known variables, we say that it is structurally globally identifiable if
\begin{equation}
    m(t,p) = m(t, \hat{p}), \forall t \geq 0 \implies \hat{p} = p 
\end{equation}

The interpretation is that the measured variables uniquely determine the values of the constant parameters for the ODE system. In such a case, it is not possible for two distinct choices of parameters to give precisely the same values for the measured variables. Note though, from
a practical perspective, that this does not disallow the possibility of $p$ and $\hat{p}$ with very different values, resulting in $m(t,p) \approx m(t, \hat{p})$. Whether a narrow range for the parameter values can be determined from an uncertain value for $m$ is the question of if the system is \emph{practically identifiable} or not. Basic global identifiability needs to be verified first though, as the practical identifiability of a system is dependent on the particular values chosen for the parameters, and the range of the variables~\cite{Practical}.
For an interesting  example where the range of
variables may play a crucial role in the 
practical identifiability, see, e.g., the very recent
work of~\cite{sauer}.

It is also possible for a system to fail to be globally identifiable, but still be what we call ``locally identifiable" meaning that no unique choice of parameters may be determined from $m(t,p)$, but that there are finitely many choices of $p$ that can be made. We find in section \ref{Numerical Results on Identifiability} that no cases of local but not global identifiability for any parameters are found for our considered systems.

For (\ref{SICRD Unscaled I to D}) we always assume that $C$ and $D$ are known and thus part of $m(t,p)$. $I$ is, by definition, not directly known, while $S$ and $R$ both require knowledge of $I$ to be known and thus these three compartments comprise $h(t,p)$. (Reported data on recoveries is available but this information is only on recoveries from confirmed cases and thus isn't actually the same quantity as $R$ in our model.) We want to guarantee that our model at least satisfies structural identifiability, meaning it is possible to estimate the parameters from incidence data alone.

\subsection{Numerical Results on Identifiability} \label{Numerical Results on Identifiability}

The precise calculation of structural identifiability results for a given ODE system is usually infeasible to perform by hand for all but the simplest of systems. There exist many software applications to
perform this type of calculation, such as the differential algebra techniques of \cite{Waterborne}. We elected to use the SIAN (Structural Identifiability ANalyser) package to test the identifiability of our model and several slight variations of the model \cite{SIAN}. The code runs a Monte Carlo algorithm to verify both local and global identifiability for an ODE system to within a high degree of certainty ($>99\%$). The following models were tested using the code.

The first model is the system of equations given in (\ref{SICRD Unscaled I to D}) without modifications. The second model matches (\ref{SICRD Unscaled I to D}) for $\dot{S},\dot{I},\dot{C}$ with the following equations for $\dot{D}$ and $\dot{R}$

\begin{eqnarray}\label{SICRD Unscaled}
\dot D&=& \frac{\beta}{T_R} C \\
\dot R&=& \frac{(1 - \alpha)}{T_L} I + \frac{1 - \beta}{T_R}C \nonumber
\end{eqnarray}

It is relevant to note that in (\ref{SICRD Unscaled I to D}) we allow deaths to occur from the unknown infected cases $I$, but that we assume the rate is the same as that for $C$, as well as allowing these deaths to be known. This is intended to reflect circumstances where a person diagnosed posthumously, or extremely close to death. Meanwhile, in (\ref{SICRD Unscaled}) we assume each fatality to occur in a case where the individual is first diagnosed. While (\ref{SICRD Unscaled I to D}) makes some stronger assumptions, it will be soon apparent that these or similar assumptions are necessary for identifiability. We thus need this for the parameters of the model to be capable of estimation from incidence data.

The third case is a slight modification of (\ref{SICRD Unscaled I to D}), now allowing $C$ and $I$ to have two distinct death rates, referred to as $\beta_1, \beta_2$ respectively. Explicitly the changed equations are 
\begin{eqnarray}\label{SICRD Unscaled 2 betas}
\dot D&=& \frac{\beta_2(1-\alpha)}{T_L} I + \frac{\beta_1}{T_R} C \\
\dot R&=& \frac{(1 - \beta_2)(1 - \alpha)}{T_L} I + \frac{1 - \beta_1}{T_R}C \nonumber
\end{eqnarray}
Finally, we consider the case of (\ref{SICRD Unscaled 2 betas}) but with the death information recorded separately, distinguishing cases from $I$ and $C$. Here we consider both to be known as the case where only $D_1$ is known was found to have only $\beta_1$ to be identifiable and the case where only $D_2$ is expected to be unrealistic.

\begin{eqnarray}\label{SICRD Unscaled D_1 D_2}
\dot D_1&=& \frac{\beta_1}{T_R} C \\
\dot D_2&=& \frac{\beta_2(1-\alpha)}{T_L} I \nonumber\\
\dot R&=& \frac{(1 - \beta_2)(1 - \alpha)}{T_L} I + \frac{1 - \beta_1}{T_R}C \nonumber
\end{eqnarray}

Each of the models, with varying assumptions on the assumed known parameters, were run as an example system through the code of \cite{SIAN}. In each case there were no instances of parameters which were locally, but not globally identifiable and thus such information was omitted. As a small note, in cases where $C(t)$ is the only known variable, $\alpha$, $\beta$ and $R_0$ are all not identifiable.  \\

{\tiny
\begin{table}
\begin{tabular}{||c | c | c | c | c||} 
 \hline
 Model & Known Variables & Known Parameters & Globally Identifiable & Not Globally Identifiable 
 \\ [0.5ex] 
 \hline\hline
 (\ref{SICRD Unscaled}) & $C(t), D(t)$ & $T_L, T_R, \beta$ & $[C(0), D(0)]$ & $[\alpha, R_0, S(0), I(0), R(0)]$ \\ 
 \hline
 (\ref{SICRD Unscaled}) & $C(t), D(t)$ & none & $[T_L, T_R, \beta, C(0), D(0)]$ & $[\alpha, R_0, S(0), I(0), R(0)]$ \\
 \hline
 (\ref{SICRD Unscaled I to D}) & $C(t), D(t)$ & none & $[T_L, T_R, \alpha, \beta, R_0, S(0), I(0), C(0), D(0)]$ & $R(0)$ \\
 \hline
 (\ref{SICRD Unscaled 2 betas}) & $C(t), D(t)$ & $T_L, T_R, \beta_1$ & $[C(0), D(0)]$ & $[I(0), S(0), \alpha, \beta_2, R_0, R(0)]$ \\
 \hline
 (\ref{SICRD Unscaled 2 betas}) & $C(t), D(t)$ & $T_L, T_R, \beta_2$ & $[\alpha, \beta_1, R_0, S(0), I(0), C(0), D(0)]$ & $R(0)$ \\
 \hline
 (\ref{SICRD Unscaled 2 betas}) & $C(t), D(t)$ & $T_L, T_R, \beta_1, \beta_2$ & $[\alpha, R_0, S(0), I(0), C(0), D(0)]$ & $R(0)$ \\
 \hline
 (\ref{SICRD Unscaled D_1 D_2}) & $C(t), D_1(t), D_2(t)$ & $T_L, T_R, \beta_1$ & $[C(0), D_1(0), D_2(0)]$ & $[\alpha, \beta_2, R_0, S(0), I(0), R(0)]$ \\ 
 \hline
 (\ref{SICRD Unscaled D_1 D_2}) & $C(t), D_1(t), D_2(t)$ & $T_L, T_R, \beta_2$ & $[\alpha, \beta_1, R_0, S(0), I(0), C(0), D_1(0), D_2(0)]$& $R(0)$ \\ 
 \hline
 (\ref{SICRD Unscaled D_1 D_2}) & $C(t), D_1(t), D_2(t)$ & $T_L, T_R, \beta_1, \beta_2$ & $[\alpha, R_0, S(0), I(0), C(0), D_1(0), D_2(0)]$ & $R(0)$ \\ [1ex]
 \hline
\end{tabular}
\caption{\label{model identifiability} Comparing Model Identifiability}
\end{table}
}

We note that in each case $R(0)$ is not identifiable, due to the fact that the current value for $R$ does not affect the rate at which any compartment is changing. It actually is able to be estimated as well, if each other variable's initial value is estimated and the total population is known. 

We see from Table 1 that under even the worst cases that $T_L$ and $T_R$ may be estimated, which aligns with the common assumption that they can be known directly from known incidence data. To estimate any other parameter requires either knowledge of the death rate for $I$, or for $I$ and $C$ to be assumed to have the same death rate. We note again that in the cases where identifiability of parameters failed that not even local identifiability held, meaning that an infinite number of parameter combinations could give rise to the exact same solution for the known variables. Thus, in order to numerically estimate the parameters we move forward with model (\ref{SICRD Unscaled I to D}) as it has the best conditions for parameter estimation, while still having reasonable underlying assumptions. A similar approach would also work for models with hospitalization data. There, hospitalizations take on a similar role to the deaths in providing indirect information on the unknown case numbers. 

As an additional remark, the properties of
model (\ref{SICRD Unscaled}) are less good
than those of (\ref{SICRD Unscaled I to D})
because for the former the time derivative
of $D$ essentially provides information for
$C$ (which is known) and hence does not
assist towards identifying $I$. The latter
identification is, apparently, possible
within (\ref{SICRD Unscaled I to D}).
In the case of (\ref{SICRD Unscaled 2 betas}),
$\beta_2$ is needed to allow for
identifiability (once again to allow 
for detection of $I$), while $\beta_1$ connected
with the confirmed cases does not suffice.
It also may not be reasonable to know $D_1$ and $D_2$ separately as generally only total death counts are in the reported information. The combination of this factor, as well as the knowledge that even with both death counts that identifiability only holds in this model when $\beta_2$ is known, which is unreasonable if it's assumed distinct from $\beta_1$, supports that model \ref{SICRD Unscaled 2 betas} is not 
as useful practically.

While having two distinct values of $\beta$ would be more realistic, we see that it creates fundamental issues with the identifiability of the model. We thus make the assumption of a single $\beta$ value for now, in order to use it as a test case for our network. 
There may be potential methods to resolve these issues while retaining identifiability, but in the present work our intention is  to keep the assumptions of the model relatively simple while testing the novel loss function introduced in section \ref{Setup of the Neural Network}. Application of the network to a more complex model is left to future work.

\section{Data Processing and Network Development}\label{Data Processing and Network Development}

\subsection{Denoising of data using a wavelet transform}\label{Denoising of data using wavelet transform}

The question of structural identifiability is one of estimation of parameters using perfect data. Real data is often highly noisy due to misreporting, late reporting, and general inconsistency concerning the gathering of information~\cite{Asymptomatic}. As pointed out in \cite{Practical}, there can exist substantial regimes in systems that are globally structurally identifiable but the ability to make useful estimates 
based on available data may be  limited. 
There may exist regimes where the sensitivity of unknown variables to known variables is high, so a slight difference in the known data may cause substantial changes in the estimates of the unknown data. There may also exist regimes where the sensitivity of the parameters to the known variables is low, so that it becomes more difficult to accurately fit values for the parameters as larger changes in parameter values may still produce a close fit to the known variables.

Before feeding our data to the neural network we have constructed, we first process the data using a wavelet transform to separate the noise of the signal from its primary features. We implement code from the PyWavelets package to perform our wavelet analysis and denoising on our signal \cite{wavelet_code}. 
More specifically, in our model, each piece of measured signal data can be mathematically expressed as
\begin{equation}\label{noisefeature}
\bf(t) = \bf_{true}(t) + \eps, \,\,\,\,\,\forall t\in [0,T]
\end{equation}
where $\bf_{true}\in \bR^{ T}$ is the true data that would have been obtained in ideal measuring conditions and $\eps$ comprises the adverse effects of the local  environment  or faulty  activity for the feature, and is referred to as “noise”. To avoid dealing with the noise in the data set, we apply a wavelet transform to denoise the data.

Wavelet theory provides a mathematical tool for hierarchically
decomposing signals and, hence, constitutes
an elegant technique for representing
signals at multiple levels of detail  \cite{Merry2005}.
In general, the wavelet transform is generated by the choice of a single “mother”
wavelet $\psi(t)$. In our implementation we used ``Symlets 5" as our choice of mother wavelet (details can be found in 
the code documentation of \cite{wavelet_code}). Wavelets at different locations and spatial scales are formed by
translating and scaling the mother wavelet. The translation and dilation of the mother wavelet are written with the operator $U(u,v)$ acting by the following \cite{cont_and_discrete_wavelet}:

$$
  U_{u,v}\psi(t) = e^{-u/2}\psi(e^{-u}t - v)
$$

Essentially we can treat the ``daughter" wavelets above, generated by rescaling and shifting the ``mother" wavelet, as analogous to the sinusoidal functions of the Fourier transform. The advantage is that the rescaling and shifting allows the wavelets to capture more local behavior, by taking wavelets with small scales $u$, and shifting across the considered time range with varying values of $v$. Potentially very robust analysis of signals may be conducted in this way, though for our purposes the goal is to use the wavelets to decompose the signal into high and low frequency components, then discard the especially high frequency components as noise.

Now, for a given signal $f$, the wavelet transform of $f$, $\Phi_{\psi}f(u,v)$ at scale $u$ and location $v$ is given by the inner product:
$$
  \Phi_{\psi}f(u,v)=\int_{-\infty}^\infty f(t) (U_{u,v}\psi(t))^*\,dt = <f, U(u,v)\psi>
$$
Here, star represents the complex conjugate.
This is known as the continuous wavelet transform, or CWT.
To ensure that the inverse CWT is well-defined, we need that 
\begin{equation}
C_\psi := \int_0^\infty \frac{|\hat{\psi}(\xi) |^2}{\xi} d\xi < \infty
\end{equation}
The hat represents the Fourier transform and this is referred to as the admissibility condition \cite{cont_and_discrete_wavelet}. One interpretation of this is that the choice of mother wavelet must have no zero frequency component, i.e. no nonzero constant component. The finiteness of this integral guarantees that the result of the CWT always still has finite $L^2$ norm. Generally once this integral is verified to be finite, the mother wavelet is rescaled by it so that the inverse  CWT may simply be defined as

$$f(t)=\int_{-\infty}^{\infty}\int_{-\infty}^{\infty} \Phi_\psi f(u,v) U(u,v)(\psi(t)) \, dudv.$$

This method of constructing the wavelet transform proceeds by producing the wavelets directly in the signal domain, through scaling and translation. When the signal frequency is higher, the wavelet base with higher time domain resolution and lower frequency resolution is used for analysis; conversely, when the signal frequency is lower, lower time domain resolution and higher frequency resolution are used. This adaptive resolution analysis performance of wavelet transform can effectively distinguish the local characteristics of the signal and the high-frequency noise, and accordingly perform the noise filtering of the signal.

\begin{figure}[ht]
  \hspace{-0.5in}
   \centering
    \includegraphics[width=6in]{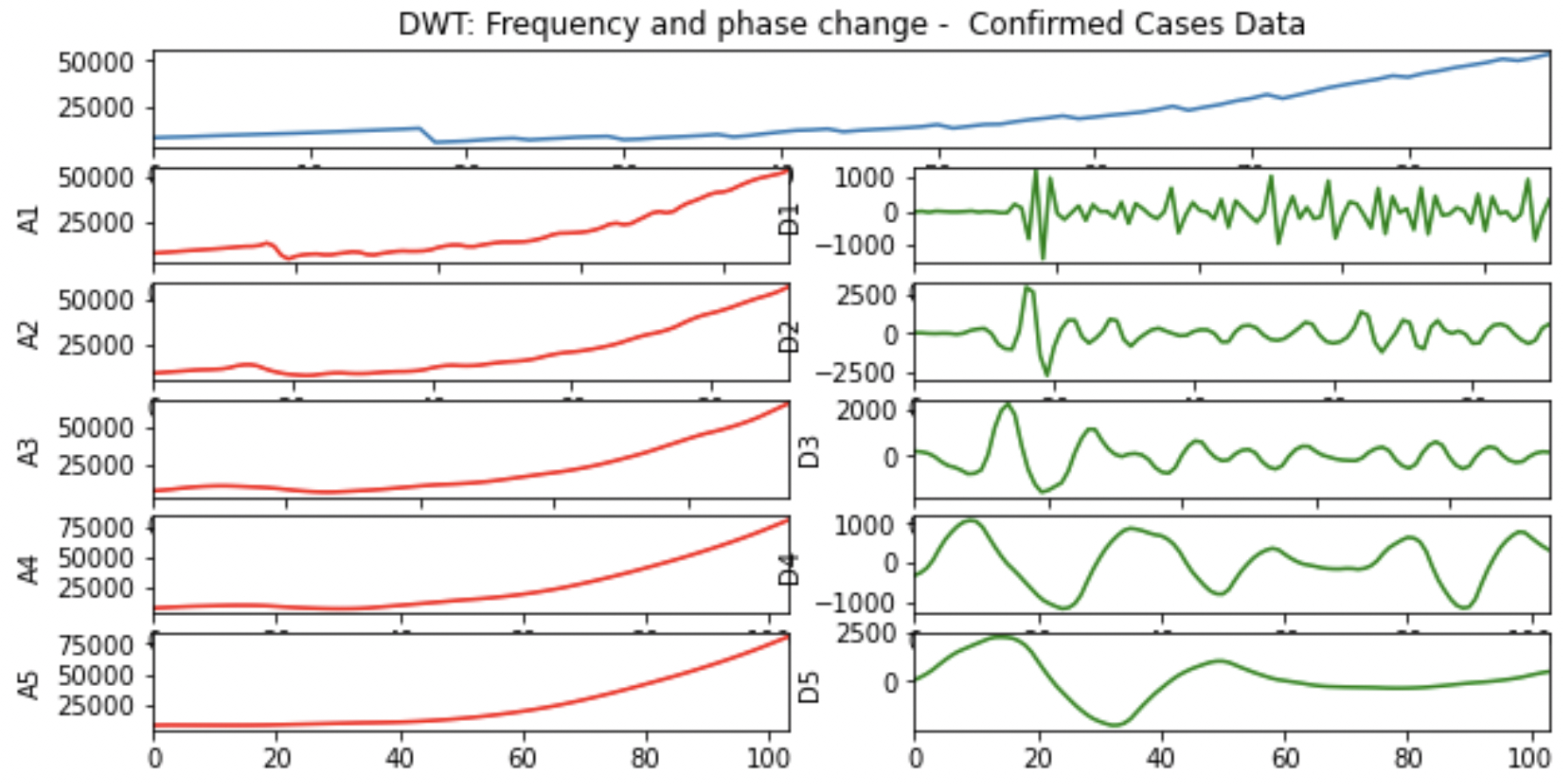}
    \caption{Wavelet decomposition of confirmed case data for Alabama starting May 1st 2020. The original time series and the subsequent five layers of decomposition into approximations ($a_i$)
    and high-frequency signal ($d_i$) are shown (see the
    text for details).}
    \label{wavelet_figure}
\end{figure}

The general steps of the wavelet transform threshold filtering method are as follows:

1) Perform the multi-scale wavelet decomposition on noisy time series signals; this process can be continued until the ``noisy" or detailed component of the signal is sufficiently low in variance (see Figure \ref{wavelet_figure}).

2) Determine a reasonable cutoff threshold and 
eliminate the high-frequency coefficients at each scale after decomposition;

3) Perform wavelet signal reconstruction from the wavelet coefficients after the zeroing process to obtain the filtered denoised signal.

The original signal $f=a_1+d_1$ can be decomposed into approximation $a_1$ as well as high-frequency signal $d_1$. Essentially we are decomposing the original signal into a sum of wavelet terms of the form $U_{u,v}\psi(t)$, with gradually decreasing values of $u$ and a suitable value of $v$. After reaching a cutoff for $u$, we reassemble the function using the approximation via lower frequency and larger scale wavelets, giving $a_1$, and the remaining portion of the function is represented as $d_1$. The process may be repeated, treating $a_1$ as the new ``original'' signal, to get the decomposition $a_1 = a_2 + d_2$, and so on.

Each time series has noticeable noise (with $C(t)$, in particular, we can see in the first image of Figure~\ref{wavelet_figure}). In addition to the removal of noise, filtering also smooths the data out as well. This smoothing doesn't substantially change the actual numerical values or trends, but it does help achieve a better fit for the model. As an ODE, the SICDR model assumes each of the variables is at least continuously differentiable, thus smoothing the inherently non-differentiable accessible data helps us perform analysis on it. This decomposition allows us to obtain a denoised signal, while still retaining information about the exact nature of the filtered noise. With the signal appropriately processed, we can now define the network intended to analyze our data. 

\subsection{Setup of the Neural Network}\label{Setup of the Neural Network}

To study the data, we implement a new ``modified" ``Physics Informed Neural Network" (PINN) to learn the values of the parameters and the unknown variable $I(t)$. The ability of neural networks to act as universal approximators is well known \cite{Multilayer} but the ability to perform accurate estimations in reasonable time frames is a central element of their increasing appeal. A variety of network structures have been developed to suit the features of particular problems. For example, recurrent neural networks like the Long Short Term Memory (LSTM) network have been used to improve pattern recognition in data that has historical dependencies~\cite{LSTM}. The PINN approach involves the changing of the loss function for the network, rather than any particular change in the network architecture itself, and in fact is compatible with a wide variety of possible architectures \cite{PINN_Survey,karnia}. The concept of the PINN is to incorporate some physical law that must be obeyed by the initial system, and to introduce an extra term into the loss function which becomes smaller the closer the network output adheres to the law \cite{PINN}. The approach was developed to estimate parameters and do forecasting in cases where data is only available at relatively sparse time intervals.

Initially, this concept was introduced in the context of nonlinear partial differential equations, though we see that our case of a nonlinear ordinary differential equation still falls into the applicable category (and other studies have also indeed used them in this context~\cite{weizhu,DINN}).  The common approach is to simply take the difference of the differential equations' sides and use that as a term to minimize in the loss function, see eq. (\ref{pinn standard loss}). This is sufficient for many applications but we found it unable to generate acceptable results in our case. This is largely due to our case not only not having extensive enough data but also having a lack of any data on some variables entirely. We also only have a single time series from the system to use in estimating the parameters, whereas in other attempts to estimate system parameters using PINN methods an entire ensemble of starting conditions and subsequent trajectories are used. For example, the DeepXDE deep learning method is used to attempt to estimate the parameters of a Lorenz system but needs to take a large collection of different starting trajectories to achieve good estimates \cite{xde_lorenz}. 

To address our case, we developed a new method, changing the structure of the loss function in \ref{loss function pieces}. The new terms still constitute the ``physical law" for our system, but are of a form where the network's gradient descent is far more stable. This novel approach, in our view, nontrivially extends PINN methods, and hence that is why we refer to it as a ``modified" PINN system. This reveals the importance of tailoring the loss function not only to each system, but to the specific circumstances of accessible variables and parameters for a given problem. The tests on artificially generated data in section \ref{Artificial Data Testing} were extremely promising, with accurate estimations of the unknown parameters, when $R_0$ and $\alpha$ were taken to be the unknowns, even with only a single time series being used to perform estimation. This is especially noteworthy as our ODE also doesn't provide any way to solve individually for the parameters by simple rearrangement, as is possible in the Lorenz system with sufficient data. The network architecture was also just the fully connected feed-forward network, the simplest type, yet it still performed well. This attests to the efficiency of this suitably chosen loss function. 

Before we precisely define our modified approach, we first review the standard PINN implementation. Generally, although many forms of PINN systems have been used, the loss function is basically of the same form \cite{PINN_Survey,karnia}. To give a precise definition for the ODE case, consider an n-dimensional ODE system defined by
\begin{equation}
    \dot{X} = f(X,p), 
\end{equation}
where $f:= f(X,p)$ is some function of all our variables $X$, and the given system parameters. The loss term $\text{Loss}_{pinn,1}$ may then be defined by.

\begin{equation}\label{pinn standard loss}
    \text{Loss}_{pinn,1}:= \overset{n}{\underset{i=1}{\sum}} \text{MSE}_t(f_i(X_{pred}) - (\dot{X}_{pred})_i).
\end{equation}
where $X_{pred}$ is the output of values from the neural network, and $\text{MSE}_t$ is the mean square error up until time $t$. This loss essentially measures how well the output of the neural network is obeying the physical laws that govern the system. Note that the estimated values of the system parameters appear in $f$ and  thus may be learned using this loss term.

The PINN loss is then combined with the standard loss function, with $X_{test}$ being the given test data:
\begin{equation}
    \text{Loss}_{mse,1} = \overset{n}{\underset{i = 1}{\sum}} \text{MSE}_t( (X_{pred})_i - (X_{test})_i).
\end{equation}
Thus, we have the total loss:
\begin{equation}
    \text{Loss}_1 =  \text{Loss}_{pinn,1} + \text{Loss}_{mse,1}
\end{equation}

The approach works, in principle, quite generally, and our initial testing essentially followed the above scheme, using the ``Disease Informed Neural Network" code as a base \cite{DINN}. Our version of $\text{Loss}_{mse, 1}$ could only include $C$ and $D$, due to these being (assumed to be) the only known variables, but otherwise the loss function was the same. It was found that the convergence was very slow in our case, with often incorrect estimations and large loss values. The existence of unknown variables creates several complications: the network has a fundamental gap in its training data, and the network is trying to estimate the unknown variable $I$ in addition to estimating the unknown parameters. The presence of so much unknown information created substantial problems in the network's ability to minimize loss for any particular term from $f_i$, without creating significant losses in another element of the equations $f_i$. The network also unevenly weighed the loss terms, as the scale of $\dot{D}$ is much smaller than $\dot{C}$.

Motivated by these deficiencies,
we developed a modified interpretation of the PINN concept in the following way. If our variables for an ODE system are given by the vector $\mathbf{X}(t)$,  we choose functions $g_i(\mathbf{X}_{pred}, p)$ (where we recall $p$ is the vector of parameters for the system) which should each be $0$ if the time series $\mathbf{X}_{pred}$ precisely obeys the ODE system. Concretely, we consider the SICRD model. For this model, $p = (T_L, T_R, \alpha, \beta, R_0)$, we formulate the functions $g_i$ from (\ref{SICRD Unscaled I to D}):

\begin{eqnarray}\label{loss function pieces}
g_1 &=& (1-\alpha)\dot C-\alpha\dot D/\beta +C/T_R \\
g_2 &=& \frac{\alpha I}{T_L} -\dot C - \frac{C}{T_R} \nonumber \\
g_3 &=& \dot I+\dot S+\dot C+\dot D/\beta \nonumber \\
g_4 &=& R_0  I S -N T_L \dot S. \nonumber \\
\end{eqnarray}

Notice that these expressions involve
suitable modifications of the dynamical
equations through (linear) combinations 
thereof.
We can then use these functions to define the term:
\begin{equation}\label{Loss_pinn definition}
    \text{Loss}_{pinn} = \sum_{i=1}^{i=4} \text{MSE}(g_i)
    \end{equation}
Then, the total loss function is given by
\begin{equation}
    \text{Loss}_{total} = \text{Loss}_{mse} + \text{Loss}_{pinn}
\end{equation}
where
\begin{equation}\label{New PINN loss}
    \text{Loss}_{mse} = MSE(C_{pred} - C_{test}) + MSE(D_{pred} - D_{test})
\end{equation}

Some issues with the performance of the code were found when trying to use ratios or log scales to try to equalize the terms in the MSE loss but ultimately the performance of the network worked well enough with the standard form. Alternative approaches with these methods of equalizing could be considered though. With these new loss functions, we were able to get the network to converge quickly and accurately when testing on artificially generated data that directly obeyed the ODE system.

Part of the motivation for this choice is that
it excludes from consideration the recovered population
for which there are always unidentifiable features (such as $R(0)$)
and focuses on the rest of the populations, including an effective
rewrite of the conservation law associated with the total population.
We also note that in our new equations we try to reduce how frequently multiple unknown quantities appear together. To illustrate this point, $g_1$ has all quantities known except for $\alpha$ if we assume $\beta$ to be known. This allows this equation to be used to learn one parameter. If $\alpha$ is in principle known from $g_1$, then all quantities except $I$ are known in $g_2$, and thus $I$ can be well estimated. Then all quantities except $\dot{S}$ are in principle known in $g_3$, letting $\dot{S}$ and $S$ itself be estimated. Finally $g_4$ can then estimate 
the basic reproduction number  $R_0$. 

Practically, this isn't precisely what is happening, as each term is being minimized simultaneously by the network, this is just meant to illustrate in principle how this structure reduces the interference of minimizing some terms while potentially causing others to grow. We also note that all of the equations are of similar order now, whereas by directly using the equations from the initial ODE the magnitude can vary drastically, as $\dot{D}$ is substantially smaller than any other derivatives appearing in the system. In that light, the present restructuring of the equations systematically builds the optimization of the system parameters.

This approach is in no way unique to our particular choice of system either. It shows that the concept of the PINN can be made so much broader than the initial obvious choice of how to incorporate information from the original set of differential equations a system is assumed to obey. Various different ODE or PDE systems could have alternatively derived formulations that can then be used to train a neural network. The most crucial point of our example is that it shows the existence of cases where the basic PINN approach is insufficient, but the modified approach is extremely effective. 

As far as the structure of the network itself
is concerned, we note that it is a simple fully connected neural network, with some minor modifications of the base code of \cite{DINN}. The hidden dimension and learning rate initially presented in \cite{DINN} created some issues, due to the original work assuming that all of the population compartments were known. This assumption allowed the original code to still converge reasonably well with a much lower learning rate, as well as a lower hidden dimension. In our implementation, we created 6 layers for our fully connected linear neural network, with hidden dimension 64. The nonlinear activation function in each layer was chosen to be the usual ReLu function, defined as: $\text{ReLu}(x)=x$, if $x>0$, and $\text{ReLu}(x)=0$ otherwise.  The learning rate was taken to be $lr= 0.0001$, while the optimizer used is Adam. We ran $12000$ epochs, to get a near perfect convergence of the estimated parameters $R_0$, $\alpha$, and the estimated unknown variable $I$ as elaborated on further in section \ref{Artificial Data Testing}. The important takeaway is that our modified PINN approach is powerful enough to achieve effective results even with one of the simplest of network architectures.

\section{Testing}
\subsection{Artificial Data Testing}\label{Artificial Data Testing}
In order to test the validity of our estimation method we first tested it on a set of artificially generated data. This was done as in the real case there is no way for us to truly know if our estimation of the parameters, or for $I(t)$ are close to the real values, meaning we can't verify the accuracy of the network beyond matching known data. We assume the real system at least roughly corresponds to the description of (\ref{noisefeature}), behaving on average as the ODE system (\ref{SICRD Unscaled I to D}) prescribes, with some random noise.

We test the neural network on artificial data with no noise, i.e., the ideal case, to verify its baseline capabilities. We generate the time series $X(t)$, a vector with each component corresponding to one of the population compartments of the SICRD model using a standard ODE solver. We then feed only the ``known" information $C(t), D(t)$, to our network with our new loss function (\ref{New PINN loss}), to generate an estimated time series. We can then measure $\text{MSE}_t(I_{pred}, I_{test})$, as well as the difference between the estimated parameters and the true values. Small MSE values suggest that so long as our base assumptions about the system are reasonable, our method accurately estimates the infected population and unknown parameters.

\begin{figure}[ht] 
    \centering
     \begin{subfigure}[b]{0.80\textwidth}
         \centering
 \includegraphics[width=\textwidth]{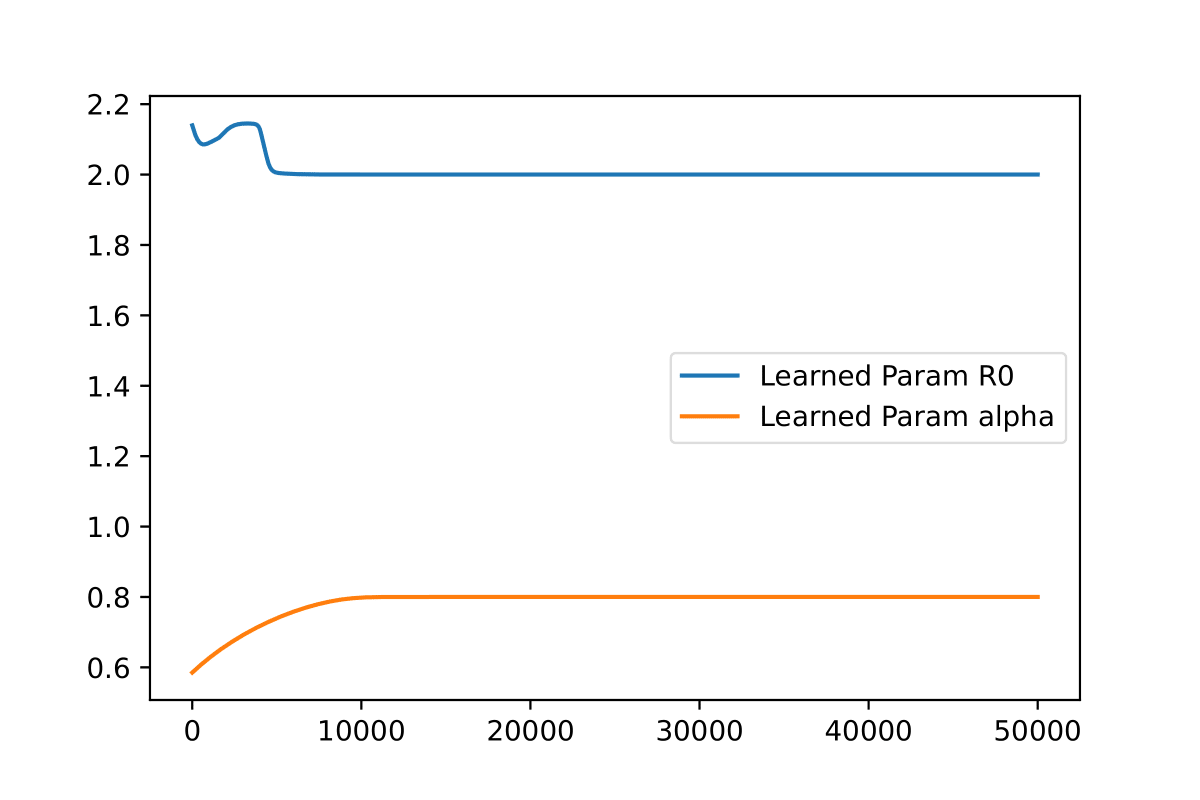}
     \end{subfigure}
 \begin{subfigure}[b]{0.80\textwidth}
         \centering
 \includegraphics[width=\textwidth]{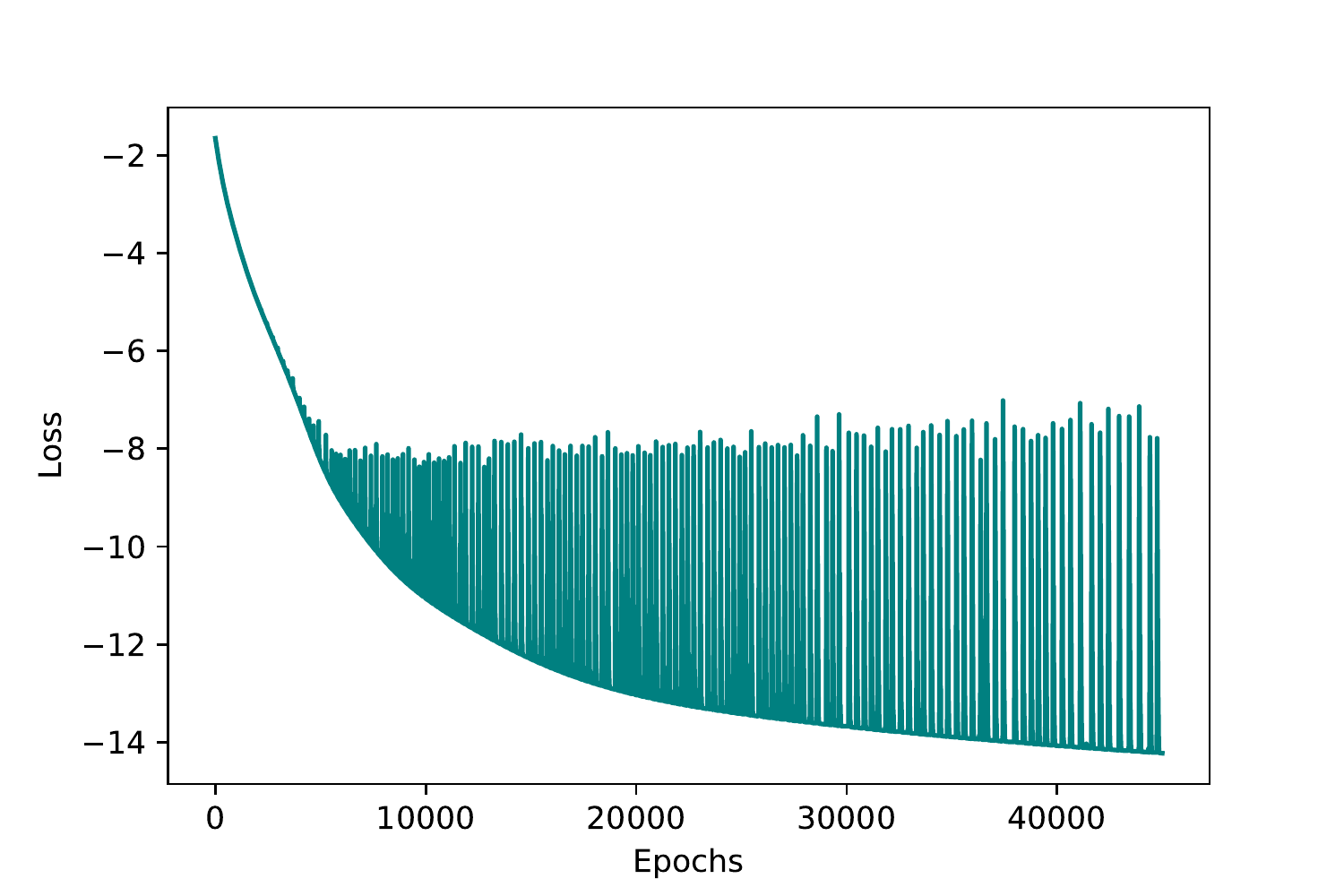}
     \end{subfigure}
    \caption{(a) The learned parameters $R_0=2$ and $\alpha=0.8$ in the  artificial data testing; (b) The log-loss curve for the training process as a function of the number of Epochs.}
    \label{artificial data figure results}
\end{figure}

For our initial test we  fixed as known the 
values of $T_L = 8.3$, $T_R = 9.2$ and $\beta = 0.05$. $R_0 = 2$ and $\alpha = 0.8$ were chosen for the unknown parameter values. We ran over a time interval of 40 days with 400 data points distributed evenly. The code was run for 50,000 epochs and executed in 3 hours. We see in figure \ref{artificial data figure results} that the correct values for the unknown parameters were converged to quite rapidly and in a stable fashion. The oscillations in the loss function  are not unexpected, as we note that the oscillations are small due to the logarithmic scale. This is essentially just the process where once the loss is small enough, the network cannot further diminish the loss without passing through regions where the loss increases temporarily as it performs gradient descent.

\begin{figure}[ht]
    \centering
    \includegraphics[width=\textwidth]{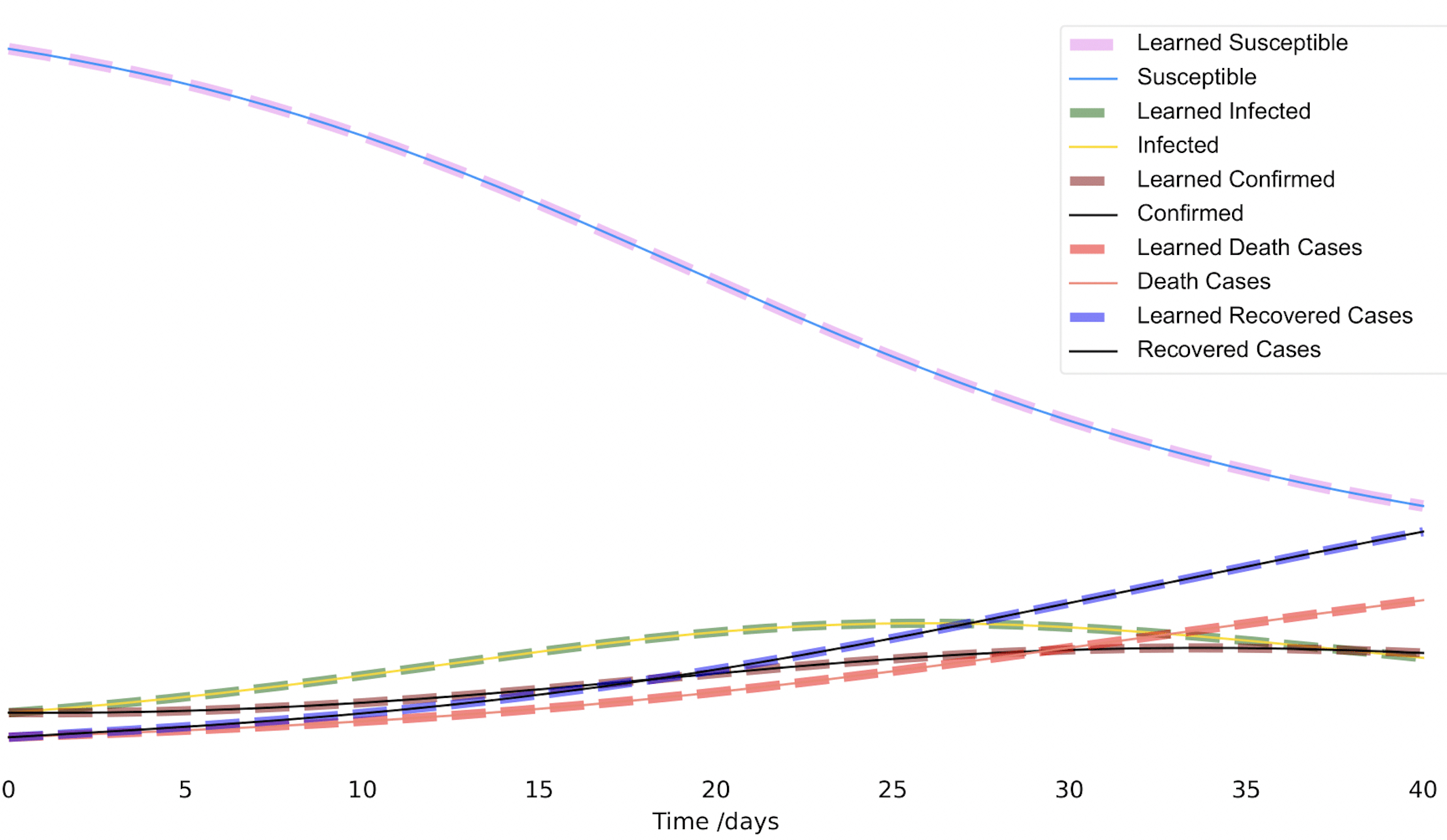}
    \caption{Estimated time series with artificial data and its comparison with the 
    artificial data of the relevant example;
    see the legend for each of the relevant
    comparisons.}
    \label{learned series artifical data}
\end{figure}

Moreover, the total MSE loss along the time series we obtain is:
$$\text{S total loss}:  0.09977164798891111;\,\,\,\,
\text{I total loss}:  0.000690453365764647;$$
$$\text{D total loss}:  0.002624943052849312;\,\,\,\,
\text{C total loss}:  0.0009282139315153407$$

Visually, we can see in figure \ref{learned series artifical data} how each learned time series (the dotted line) overlaps almost precisely with the true time-series. Clearly, in this 
artificial data example, the DINN performs
very adequately in identifying both the parameters
and the unavailable time series information.

\subsection{Real Data Testing}\label{real data testing}

After verifying the effectiveness of the network at testing on artificial data, the next step was to perform testing on real data to measure its effectiveness. Only $C$ and $D$ are known variables with the real data though, so it is not possible to evaluate the network based on its ability to estimate the unknown variables. It's still possible to at least measure how well the network is able to fit the available data while generating estimated time series for the unknown variables. The network was trained on reported data from Alabama for 90 days starting on May 1st, 2020.

\begin{figure}[ht] 
    \centering
     \begin{subfigure}[b]{0.60\textwidth}
         \centering
 \includegraphics[width=\textwidth]{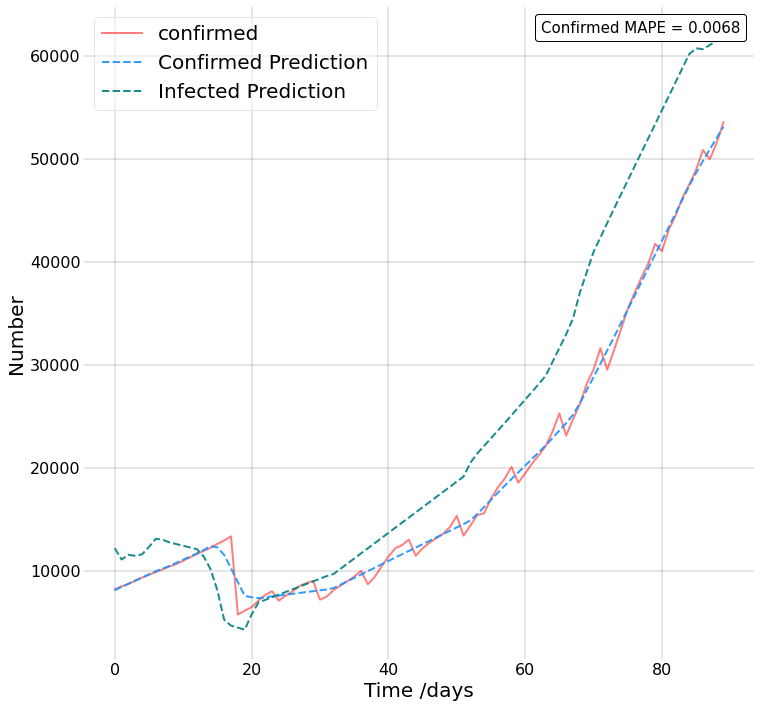}
     \end{subfigure}
 \begin{subfigure}[b]{0.60\textwidth}
         \centering
 \includegraphics[width=\textwidth]{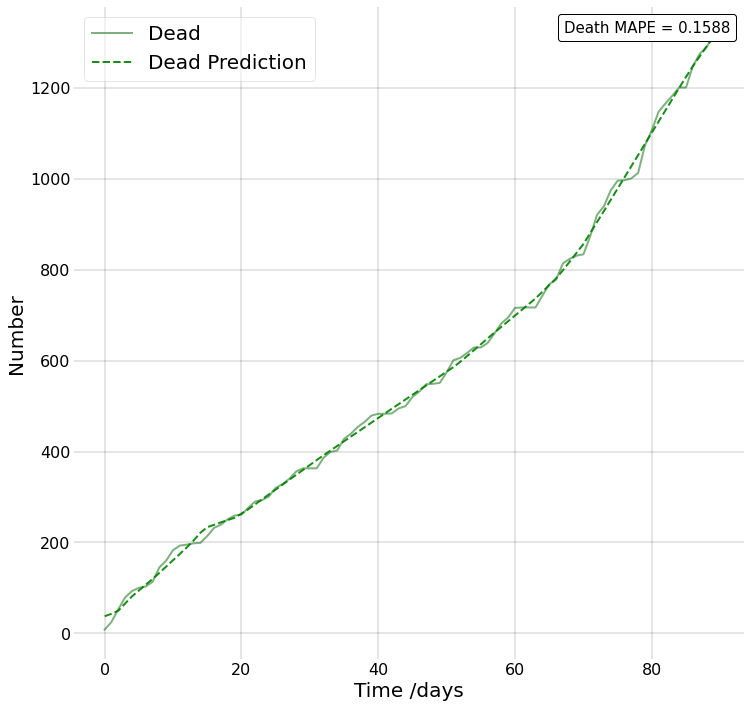}
     \end{subfigure}
     \caption{(a) The estimated values for infected and confirmed cases, compared with the actual confirmed count; (b) The estimated values for death cases, compared with the actual death count. In each case the
     Mean Absolute Percentage Error (MAPE) is included.}
     \label{real data figure results}
\end{figure}

We see that the network was able to generate very close fits to the known variables $C$ and $D$ over the 90 day time span, despite the relatively noisy data. We do not train over longer intervals of time due to the eventual buildup of errors from the assumption of constant parameter values. A potential approach using time varying parameters is proposed in \ref{time dependent parameters}, but initial testing is constrained to constant parameter values in order to perform 
our initial verification.

Thus so long as the model is a reasonable fit, the estimates for $I(t)$ and the unknown parameters should be relatively accurate in their
determination. Note that adaptation of the network approach could easily be applied to other models as well, though new functions for (\ref{loss function pieces}) would have to be derived.

\subsection{Learning the time-dependent parameters}\label{time dependent parameters}

The assumption that the parameters in equation (\ref{SICRD Unscaled I to D}) are constant (i.e., time-independent) is a highly restrictive, and unrealistic one for long enough time scales. In the real world scenario, many of these parameters are changing as time passes, due to factors such as mutation of the virus and shifting government policy;
the latter is well-known to modify social interactions in the
case of non-pharmacological interventions and hence affect factors
such as $R_0$~\cite{non-pharma_intervention,cuevas2021}. In this case we update the model in (\ref{SICRD Unscaled I to D}) to replace $\alpha, R_0, \beta$ with $\alpha(t), R_0(t), \beta(t)$, while $T_L$ and $T_R$ are left fixed as they are relatively well known, stable parameters.
In the case of mutations, it is not 
unreasonable to expect variations of these
parameters too, but for the cases under
consideration, we expect such variations
to be small and anyway secondary to the above
effects.

We find that our modified PINN is able to effectively and efficiently minimize loss over a scale of about 60-90 days. Due to efficiency of the code we can reasonably allow the parameters to vary by simply allowing a ``rolling window" of constant parameter estimations. To illustrate, we consider $\alpha(t)$. We first fix a positive integer constant, $\Delta_t$, in our case 60 or 90. We then allow $\alpha(t)$ to be equal to the estimated constant value for the parameter over the interval $[t, t+\Delta_t]$.

This gives a simple, but robust method of more precisely estimating the variation of the parameters over time. Of particular note is that we make no specific assumptions about the type of function for each parameter. The network runs efficiently enough to allow a totally general form for the parameter values, allowing the capture of potentially unusual behavior that might not be seen if a particular form was assumed. To predict later parameter values we assume the parameters have been determined on the interval $[0,T]$, using data on the interval $[0,T+\Delta_t]$. To determine $\alpha(T+1)$ we simply assume that
\begin{equation}
    \alpha(t)|_{[T,T+1)} = \alpha(T).
\end{equation}
The same assumption is used for all other varying parameters, and these values may then be used to generate the population compartment values for the next time step. The neural network's training process is then repeated to learn the value of $\alpha(T+1)$, using the predicted component values for the future as new assumed training data. Varying values for the parameters were found using available data in Alabama from May 1st 2020 to Dec 31st 2020, with a rolling window of 90 days. 

\begin{figure}[!ht]
    \includegraphics[width=.68\textwidth]{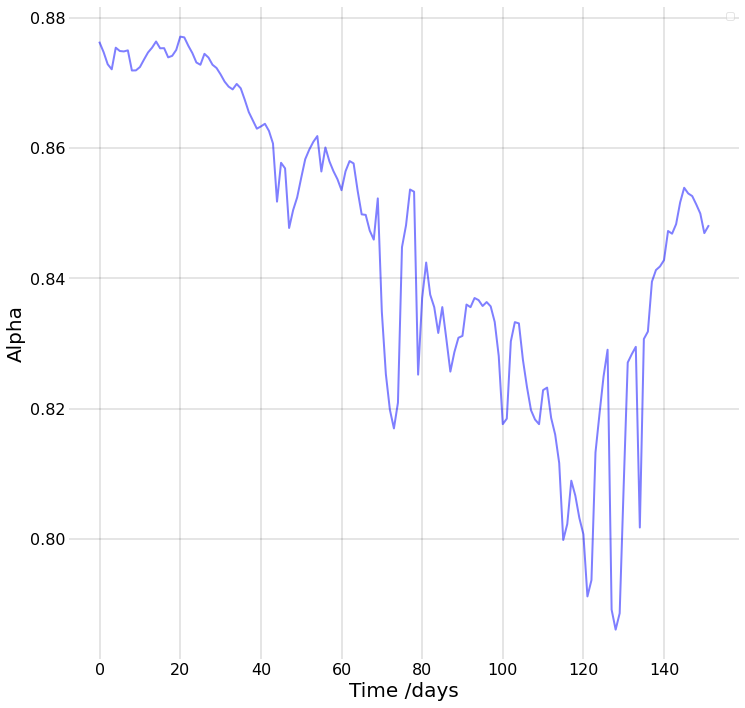}
    \includegraphics[width=.68\textwidth]{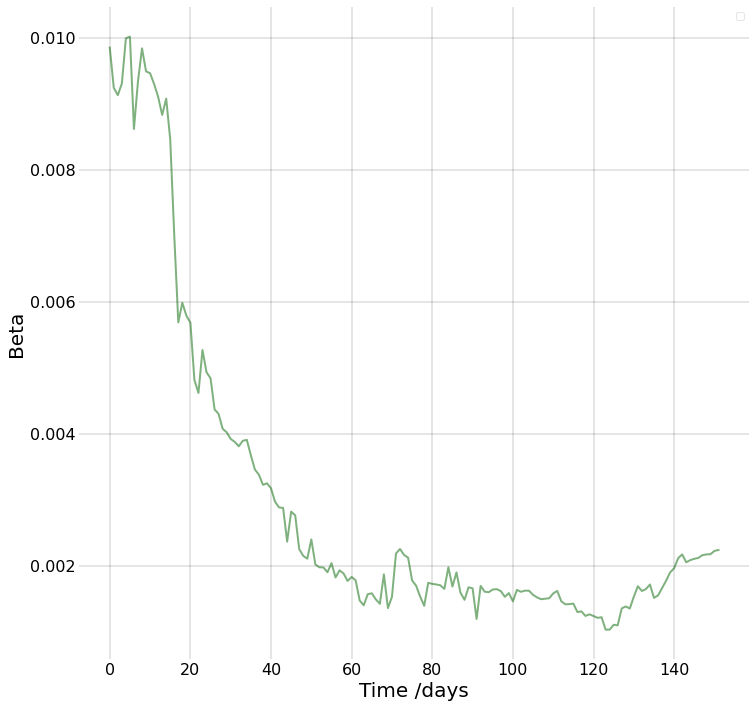}
    \caption{Learned values of the time dependent parameters $\alpha$ and $\beta$ and $R_0$ as a function of time from May 1, 2020
    for Alabama.}
    \label{varying parameters}
\end{figure}

We see from our estimates of the time varying parameters in Figure~\ref{varying parameters} that there do appear to be meaningful variations in their values. If the true values were constant, or close to constant, we would not expect to see such substantial variation in the graphs, so it seems that realistically the parameters should definitely be assumed to vary. 
Note that $\alpha$ does vary to some degree, while $\beta$ has a drastic drop over the time frame. The value of $R_0$ seems relatively stable, suggesting that there weren't substantial social distancing efforts implemented. The confirmation rate $\alpha$ seems to indicate a difficulty in keeping testing up with infections at first, though some eventual adaptation
takes place. The highly ``noisy" look of $\alpha$ may likely be due to artifacts of how confirmed cases are reported. The most substantial variation though is the drop in 
$\beta$, which may reflect an increased awareness of how to treat severe cases of COVID-19. Care has to be taken in interpreting the values though, as extraneous factors such as reclassification of what is considered a COVID-19 fatality may muddle the available data. 
The results 
are suggestive of the viability of the use of this network approach on systems with time varying parameters. 
In the case example that now follows, we restricted ourselves to a 90 day time window with the assumption of fixed parameters in section \ref{Ranking States by Testing} in order to prevent some unforeseen issues. 

\begin{figure}[!ht]
    \includegraphics[width=.68\textwidth]{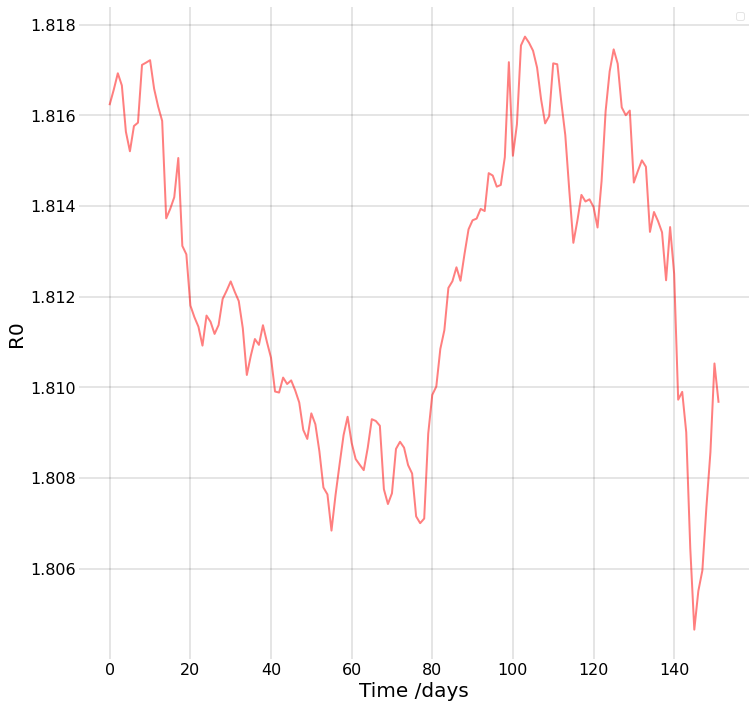}
    \caption{Learned values of the time dependent parameter $R_0$}
    \label{varying parameters 2}
\end{figure}

\section{Ranking States by Testing}\label{Ranking States by Testing}

Now that we have a network that is able to, with reasonable effectiveness and efficiency, estimate these unknown quantities, we consider a concrete question this information can help us answer, ``Which states have the worst overall rates of testing?" This is an important piece of information for federal policymakers, as it determines which areas would be in most dire need of additional federal aid in the form of (the limited) testing supplies. The sense of ``worst testing rate" can be quantified by finding which states have the highest ratio of $I$ to $C$, averaged over a given time interval. Explicitly, we select
as our corresponding diagnostic:
\begin{equation}
    A(t) = \frac{1}{t} \sum_{k=1}^t \frac{I(k)}{C(k)}
\end{equation}
The importance of $A(t)$ is that it lets us estimate which areas of the country are struggling the most to efficiently test their populations, relative to the prevalence of the disease. The idea of this particular metric is that it gives more specific information than just simply taking the estimation for the unknown infected population and dividing it by the population. Taking the infected population alone might just give information on locations that happen to have an exceptionally high number of infections occurring. The selected metric more directly identifies areas (in our case: states) where information is relatively scarce, which could be vital for identifying population centers that may not have a high number of infections currently, but are highly vulnerable to potential outbreaks due to poor testing infrastructure. The information from our direct estimates for $I(t)$ as well as for $A(t)$ together give a more robust measure of the current testing situation.

\begin{figure}[!ht]
    \includegraphics[width=.68\textwidth]{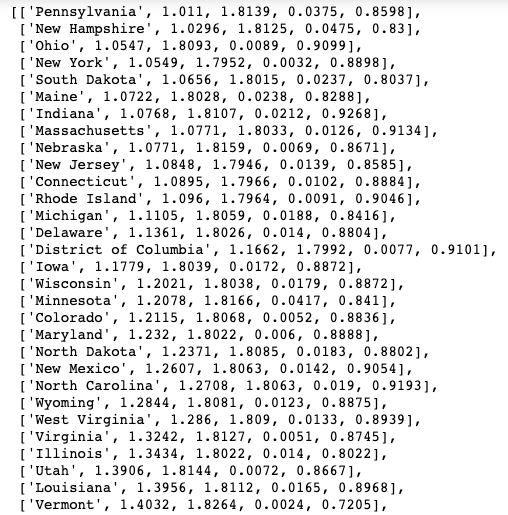}
    \includegraphics[width=.68\textwidth]{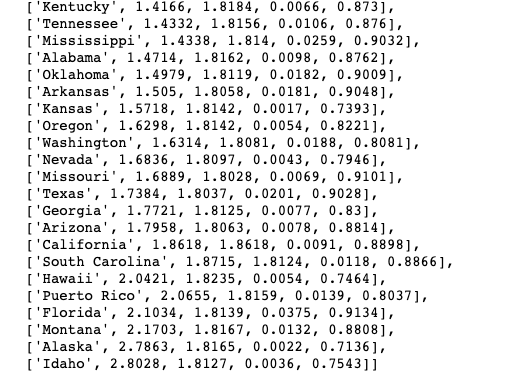}
    \caption{States organized by ranking values estimated over 90 days starting at May 1st, 2020. First column is $A(t)$, second is $R_0$, third is $\beta$ and the last is $\alpha$.}
    \label{ranking table}
\end{figure}

The importance of early intervention in the spreading of COVID-19 is already known \cite{ZhangDong20} so the potential for this metric to provide an early warning is 
highly valuable. It should be noted that while there are limitations in estimating $I(t)$, as long as the network is able to distinguish population centers of major need from ones of minor need, then it is still useful as a tool for policy making. The values of $A(t)$ over a 90 day interval starting on May 1st 2020, estimated by our network, are given in table (\ref{ranking table}).

At the highest extreme we have Idaho with a roughly 2.8 to one ratio for infected to confirmed. We note that while the timescale for this type of statistics was early in the pandemic, a state's ranking can be found
to have correlations with more long-term effects. During the tested period, Idaho would neglect to implement mask mandates in the summer of 2020 \cite{Idaho_Mask_Mandate}, later facing extreme overburden of its medical system by December 2020 \cite{Overtaxed_Idaho}, and further on having the lowest vaccination level later in September 2021 \cite{Idaho_Vaccination}, features that appear to corroborate our findings. 

Pennsylvania had the lowest ranking, with a ratio of 1.011, just barely more cases of unknown infections than confirmed cases estimated. Pennsylvania had already instituted some mask mandates in April 2020 \cite{Pennsylvania_mask_april}, and these mandates were substantially expanded in July 2020 \cite{Pennsylvania_mask_july}. Universities within the state also began declaring classes would be moved online for the fall semester \cite{Pennsylvania_september_school}. 

While in the present discussion we have focused
on table \ref{ranking table}'s extremes, a closer inspection suggests that all
the highest ranking states may be the ones that one might expect
(including, e.g., New York and New England states) and similar
conclusions could be drawn for the some of the (predominantly southern)
states at the bottom of the table.

Overall, the existence of such a substantial disparity between states supports the usefulness of this type of analysis. Indeed, it is clear that there are meaningfully different levels of uncertainty in the known infected population from state to state. Realistically there are many factors that may interfere with an accurate picture of this estimate. Incorrectly attributed deaths, inconsistency with reporting, and other issues may cause errors beyond simple noise. As long as we are capable of accurately distinguishing the states with highest values for $A(t)$ then we still have useful information, and the disparities are wide enough that even with moderate error margins it is still possible to create such a categorization.

\section{Conclusions and future work}

In the present work we presented a relatively simple model for the time-evolution
of COVID-19, motivated by features of the disease and the nature
of available data. The model allowed the distinguishing of confirmed and unconfirmed cases, as well as a recording of the number of fatalities due to the disease. Upon verification of the identifiability of the 
unknown quantities within the model (based on the practically available data), we were able to use our modified PINN approach to analyze available data, with a special loss function tailored to our circumstances. There are, of course, many simplifications made throughout our work. The model could always be further separated into more compartments, to represent more distinct population compartments relevant to the spread of the disease. Compartments for exposed, though not yet infectious individuals, explicitly asymptomatic individuals, hospitalized individuals, and more could be included and we discussed some
principles on the basis of which such considerations could be explored. Allowances for individuals to lose immunity after recovery, and for individuals who have been confirmed to infect others, may also be considered, as well as (more recently) the possibility to account for vaccination. 

The intention of the model was to study the development of the disease on relatively short time frames, mitigating the effect of immunity loss as individuals would most likely still have immunity. The analysis focused on the early stages of the pandemic, and over this time frame the error caused by not accounting for individuals becoming susceptible again after infection is practically negligible. Instances of confirmed individuals spreading infection are assumed to be an exception as most individuals followed appropriate quarantining procedures on a positive test result.

Ultimately, the simplifications made for the model, as well as for the actual architecture of the neural network, were taken so that testing with the novel loss function,  as well as usage of the wavelet transform to process the data, could be performed with minimal extraneous complications. Future work could naturally focus on expanding both the model's complexity as well as incorporating more nuanced network structures. The work of \cite{GNN_Germany} on graph neural networks gives a very promising direction for future work in both directions, with a natural extension of the model to incorporate human mobility, while also having a neural network that is well suited to the study of such a system. 
Indeed, the study of such metapopulation models of wide appeal
within the modeling of COVID-19~\cite{ZhangDong20,review_meta},
in conjunction with some of the technical approaches and methodologies
presented herein constitutes a promising direction for future study.

\newpage

\bibliographystyle{plain}
\bibliography{refs}

\end{document}